\documentclass[aps,prd,preprintnumbers,superscriptaddress,nofootinbib,twocolumn]{revtex4-1}

\usepackage[dvipdfmx]{graphicx}
\usepackage{xcolor}
\usepackage{bm,latexsym,amsmath,amssymb,amsfonts,mathrsfs}
\usepackage[colorlinks=true,urlcolor=teal,citecolor=orange,linkcolor=violet]{hyperref}

\newcommand*{\ba}{\begin{eqnarray}}
\newcommand*{\ea}{\end{eqnarray}}
\newcommand*{\bb}{\begin{framed}}
\newcommand*{\eb}{\end{framed}}

\newcommand*{\mpl}{M_{\rm Pl}}
\newcommand{\simgt}{\lower.5ex\hbox{$\; \buildrel > \over \sim \;$}}
\newcommand{\simlt}{\lower.5ex\hbox{$\; \buildrel < \over \sim \;$}}

\newcommand*{\p}{\partial}

\newcommand*{\calC}{{\cal C}}

\newcommand*{\calE}{{\cal E}}

\newcommand*{\calM}{{\cal M}}

\newcommand*{\calO}{{\cal O}}
\newcommand*{\calP}{{\cal P}}

\newcommand*{\calZ}{{\cal Z}}

\newcommand*{\mgw}{M_{\rm GW}}

\newcommand*{\del}{\partial}

\def\({\biggl(}
\def\){\biggr)}
\def\[{\biggl[}
\def\]{\biggr]}

\newcommand{\nn}{\nonumber \\}

\begin{document}
\title{Viable massive gravity without nonlinear screening}

\author{Yusuke Manita}
\email{manita@tap.scphys.kyoto-u.ac.jp}
\affiliation{Department of Physics, Kyoto University, Kyoto 606-8502, Japan}

\author{Sirachak Panpanich}
\email{sirachakp@aoni.waseda.jp}
\affiliation{Department of Pure and Applied Physics, Graduate School of Advanced Science and Engineering, Waseda University, Okubo 3-4-1, Shinjuku, Tokyo 169-8555, Japan}

\author{Rampei Kimura}
\email{rampei@aoni.waseda.jp}
\affiliation{Waseda Institute for Advanced Study, Waseda University,1-6-1 Nishi-Waseda, Shinjuku, Tokyo 169-8050, Japan}

\preprint{KUNS-2951}

\date{\today}

\begin{abstract} 
    We study nonlinear effects of perturbations around a cosmological background in projected massive gravity, which admits self–accelerating solutions in an open FLRW universe. 
    Using the zero-curvature scaling limit,
    we derive nonlinear equations containing all the relevant terms on subhorizon scales. We find that the solution for a scalar graviton vanishes completely for all scales, which agrees with the linear perturbation analysis in the previous study. In addition, the effects on the gravitational potential due to the next order perturbation are strongly suppressed within the horizon. Therefore, a screening mechanism is no longer needed for consistency with solar-system experiments in the projected massive gravity.
\end{abstract}

\maketitle

\section{Introduction}
In the last decade there has been considerable progress in the discovery and development of ghost-free massive gravity, so called de Rham-Gabadadze-Tolley (dRGT) massive gravity~\cite{deRham:2010ik,deRham:2010kj}. The theory is free from the Boulware-Deser (BD) ghost~\cite{Boulware:1972yco}, typically appearing due to the breaking of general covariance, it thus has only $5$ degrees of freedom ($2$ tensor $+$ $2$ vector $+$ $1$ scalar modes). 
Although the dRGT massive gravity admits an open-Friedmann-Lema\^{\i}tre-Robertson-Walker (FLRW) universe with the effective cosmological constant driven by a mass term, it unfortunately suffers from nonlinear instabilities~\cite{DeFelice:2012mx} associated with a strong coupling of the scalar and vector graviton~\cite{Gumrukcuoglu:2011ew}. 
For this reason, the dRGT model is not a viable theory of massive gravity, and various extension has been intensively investigated in the context of new derivative interactions~\cite{Hinterbichler:2013eza,Kimura:2013ika,deRham:2013tfa,Gao:2014jja,Folkerts:2011ev}, adding an extra scalar degree of freedom~\cite{Huang:2013mha,DAmico:2012hia}, a minimal theory of massive gravity~\cite{DeFelice:2015hla,DeFelice:2020ecp,DeFelice:2017wel,DeFelice:2017rli} and bigravity~\cite{Hassan:2011zd}. 

One interesting and minimal way to extend the dRGT massive gravity without introducing an extra degree of freedom is to break the translation invariance of the St\"{u}ckelberg fields while preserving the global Lorentz invariance, and it allows to include new Lorentz-invariants in the Lagrangian. The simplest extension where the constant parameters in the potential of the dRGT theory are promoted to be arbitrary functions of the Lorentz-invariant is called generalized massive gravity (GMG) theory~\cite{DeRham:2014wnv,deRham:2014gla}. Furthermore, with an idea of the degeneracy of the St\"{u}ckelberg kinetic term, the breaking of the translational invariance can introduce a nonminimal coupling of the scalar curvature to the Lorentz-invariant quantity of the St\"{u}ckelberg field and two different classes of graviton mass terms without invoking the BD ghost \cite{Gumrukcuoglu:2021gua}.  
The first class is an straightforward extension of the dRGT theory (as well as GMG) obtained by applying the conformal and disformal deformations of the fiducial metric. In spite of the failure of the dRGT theory, the GMG theory has self-accelerating solutions which are free from any instabilities on an open-FLRW background \cite{Kenna-Allison:2019tbu}. 
The second class is called projected massive gravity (PMG) theory. The mass term is composed of a projection tensor which eliminates one of St\"{u}ckelberg fields and manifestly prevents the appearance of the BD ghost. The PMG theory similarly admits a perturbatively stable open-FLRW solution \cite{Gumrukcuoglu:2020utx}.

In massive gravity, the scalar mode typically induces a fifth force comparable to the Newtonian force. In the case of the GMG theory, this can be seen, for example, in the effective gravitational coupling in the evolution equation for the matter perturbation~\cite{Kenna-Allison:2020egn}. 
Although modified gravity theories requires a screening mechanism to hide the fifth force in order to satisfy various observations on small scales such as solar-system~\cite{Will:2014kxa}, massive gravity originally possess the Vainshtein mechanism \cite{Vainshtein:1972sx} as a solution of the vDVZ discontinuity \cite{vanDam:1970vg,Zakharov:1970cc}. This can be easily seen in the effective Lagrangian obtained by taking the decoupling limit of the dRGT theory~\cite{deRham:2010ik}, where the galileon interactions play an important role, and the additional force due to the scalar graviton can be strongly suppressed inside a characteristic scale called the Vainshtein radius \cite{Babichev:2009us,Berezhiani:2013dw}. In fact, the Vainshtein mechanism operates in the case of the GMG theory~\cite{Gumrukcuoglu:2021gua}. 

On the other hand, in the PMG theory, the linear perturbation analysis around a cosmological background shows the fifth force associated with the scalar graviton vanishes at the leading order in the subhorizon expansion, i.e.,  the absence of the vDVZ discontinuity~\cite{Manita:2021qun}. This intuitively implies that the Vainshtein mechanism is not necessary in the PMG theory. However, galileon-like interactions which potentially dominate at small scales should be present in nonlinear equations as in the GMG case. Thus it is not confirmed yet whether the fifth force comparable to the Newtonian one is absent even at small scales. To this end, we, in the present paper, investigate nonlinear solution of the scalar graviton by taking into account the cosmological background in order to check the consistency with the local experiments. 

The paper is organized as follows. In Sec. \ref{PMG}, we introduce action and basic equations of the theory. We study 
background and linear perturbation equations in the zero-curvature limit, and quasi-static limit in Sec. \ref{cospert}. In Sec. \ref{spherical}, we investigate nonlinear perturbations in a spherically symmetric setup. Lastly, Sec. \ref{discussion} is devoted to discussion and conclusion. 

In the following, we use the notation $c=1$, and $\eta_{\mu\nu}=\rm{diag.}(-1,1,1,1)$. With the Newton constant $G_{N}$, the Planck mass is defined by $\mpl:=(8\pi G_{N})^{-2}$.

\section{Projected massive gravity}
\label{PMG}

In this section, we briefly introduce the ghost-free theory of projected massive gravity (PMG)
\cite{Gumrukcuoglu:2020utx}.
The mass term in the PMG theory is defined through the following fiducial metric:  
\begin{align}
    \bar{f}_{\mu \nu}:=P_{a b} \partial_{\mu} \phi^{a} \partial_{\nu} \phi^{b}\,.
\end{align}
Here, the four scalar fields $\phi^a$ are the so-called St\"{u}ckelberg fields, and $P_{ab}$ is the projection operator on the field space defined by
\begin{align}
    P_{ab} = \eta_{ab}-\frac{\phi_a\phi_b}{X}
    \,,
    \label{projection}
\end{align}
with $X:=\eta_{ab}\phi^{a} \phi^{b}$, 
which projects onto surfaces in the field space of the St\"{u}ckelberg fields $\phi^a$. With the building block matrix of the mass term, $Z^{\mu}{}_{\nu}:= \left(g^{-1} \bar{f}\right)^{\mu}{}_{\nu}$, the action of PMG is given by
\begin{align}
    S &= \frac{\mpl^2}{2}\int d^{4} x \sqrt{-g} \Bigg[ G R
    +\frac{3G_X^2}{2G}\partial_\mu X \partial^\mu X
    \nn
    &+m^{2} U\left(X,[Z],\left[Z^{2}\right],\left[Z^{3}\right]\right)\Bigg]+S_{\mathrm{m}}\left[g_{\mu \nu}, \psi\right]
    \,,
    \label{action}
\end{align}
where $G$ is the arbitrary dimensionless function of $X$, $G_X := \partial G / \partial X$, $m$ is a constant mass parameter, $U$ is the arbitrary dimensionless function of $X, [Z], [Z^2]$, and $[Z^3]$, and $S_{\rm m}$ is the action of the matter field $\psi$ which minimally couples to the physical metric $g_{\mu\nu}$. The existence of $X$-dependence in the action \eqref{action} as well as the projection tensor \eqref{projection} manifestly break the translation invariance of the St\"{u}ckelberg fields while the global Lorentz invariance is retained. In other words, thanks to the violation of the internal translational symmetry, PMG generally includes the non-minimal coupling of $\phi^a$ to the Ricci scalar. Despite this fact, stable cosmological solutions have not been yet found in the case of $G \neq 1$~\cite{Manita:2021qun}. For this reason, we focus on the case with $G=1$ in the following part of this paper. 

The modified Einstein equation is given by varying the action with respect to $g^{\mu\nu}$,
\begin{align}
    \mpl^{2} G_{\mu\nu} = T^{(\rm mass)}_{\mu\nu}+T_{\mu\nu}^{(\rm m)}\,,
    \label{eq:eom}
\end{align}
where $T^{(\rm mass)}_{\mu\nu}$ is the effective energy-momentum tensor of the graviton mass term that is defined by
\begin{align}
    T^{(\rm mass)}_{\mu\nu} &= m^2 \mpl^2 \Big(
    {1 \over 2} g_{\mu\nu} U - U_{[Z]} {\bar f}_{\mu\nu} 
    -2 U_{[Z^2]} Z^{\rho}_{~(\mu} {\bar f}_{\nu)\rho} 
    \nn
    &-3 U_{[Z^3]} Z^{\rho}_{~\sigma} Z^{\sigma}_{~(\mu} {\bar f}_{\nu)\rho} 
    \Big) \,,
\end{align}
and $T^{(\rm m)}_{\mu\nu}$ is an energy-momentum tensor for the matter field.
We assume a non-relativistic perfect fluid for the matter field,
\begin{align}
    T^{(\rm m)}_{\mu\nu} = \rho u_{\mu} u_{\nu}
    \,,
\end{align}
where $\rho$ is the energy density and $u^{\mu}$ is a four-velocity of the matter field. Requiring the conservation of the energy-momentum tensor, 
\begin{align}
    \nabla^\mu T^{(\rm m)}_{\mu\nu} = 0\,,
    \label{eq:matter_equation}
\end{align}
the divergence of the Einstein equation~\eqref{eq:eom} gives
\begin{align}
    \nabla^\mu T^{(\rm mass)}_{\mu\nu}=0\,.
    \label{eq:Stuckelberg_equation}
\end{align}
This conservation equation of the effective energy-momentum tensor of the graviton mass provides the equation of motion of the St\"{u}ckelberg fields, which can be obtained by varying the action with respect to $\phi^a$.

\section{Cosmological perturbation}
\label{cospert}
In this section, we revisit linear perturbations around a cosmological background in PMG. We introduce the zero-curvature scaling limit for simple calculations.

\subsection{zero-curvature limit}
In the following, we consider a homogeneous and isotropic space-time with positive spatial curvature, i.e., an open ($\kappa>0$) FLRW, given by
\begin{align}
    g_{\mu\nu} dx^{\mu} dx^{\nu} &= -dt^2 + a(t)^2 \Omega_{ij}dx^i dx^j
    \,,
    \end{align}
where
\begin{align}
    \Omega_{ij} &= \delta_{ij}-\frac{\kappa\,x^ix^j}{1+\kappa\,x^kx^k}
    \,.
    \label{metric}
\end{align}
The unique configuration for St\"{u}ckelberg fields to preserve homogeneity and isotropy is provided with~\cite{Gumrukcuoglu:2011ew}
\begin{align}
    { \phi}^0 &= f(t)\,\sqrt{1+\kappa(x^2+y^2+z^2)}\,,
	\nn
	{\phi}^i &= f(t)\,\sqrt{\kappa}\,x^i
	\,,\label{eq:Stuckelberg}
\end{align}
and the ${\bar f}$-metric is then given by
\begin{align}
    \bar{f}_{\mu\nu}dx^\mu dx^\nu&=\kappa f^2\Omega_{ij}dx^idx^j
    \,.
\end{align}
Throughout this present paper, for simplicity, we use the zero-curvature scaling limit, which is first introduced in~\cite{deRham:2014gla}. 
We begin with a redefinition of the function $f(t)$ in \eqref{eq:Stuckelberg} as
\begin{align}
    f(t) = \frac{\alpha}{\sqrt{\kappa}} + \chi(t)\,,   
    \label{scaling}
\end{align}
where $\alpha$ is a constant. Taking the $\kappa\to 0$ limit after the redefinition \eqref{scaling},
the line elements for the $g$-metric and the $\bar{f}$-metric reduce to
\begin{align}
    \lim _{\kappa \to 0} {g}_{\mu\nu} dx^{\mu} dx^{\nu} &= -dt^2 + a(t)^2 \delta_{ij} d x^i d x^j
    \,,\label{eq:g_zcsl}\\
    \lim _{\kappa \to 0} {\bar{f}}_{\mu\nu} dx^{\mu} dx^{\nu} &= \alpha^2 \delta_{ij} d x^i d x^j
    \,.\label{eq:f_zcsl}
\end{align}
On the other hand, under the redefinition \eqref{scaling}, the Lorentz-invariant quantity $X$ becomes
\begin{align}
    X=-\frac{\alpha^{2}}{\kappa}-\frac{2 \alpha}{\sqrt{\kappa}} \chi(t)
    -\chi(t)^2
    \,,
\end{align}
and thus diverges in the limit of $\kappa \to 0$. To avoid this problem, we redefine $X$ as
\begin{align}
    \tilde{X} := -\frac{\sqrt{\kappa}}{2\alpha}\left( X + \frac{\alpha^2}{\kappa} \right) 
    \,,
    \label{newX}
\end{align}
and then $\tilde{X}$ converges to a finite value since $\lim_{\kappa \to 0} \tilde{X} = \chi (t)$. The mass potential $U$ is an arbitrary function of $X$ and the trace of $Z^n$, and we can thus redefine it as $\tilde{U}(\tilde{X},[Z],[Z^2],[Z^3]) := U(X,[Z],[Z^2],[Z^3])$. 
Therefore, introducing the redefinitions of \eqref{scaling} and \eqref{newX}, the action \eqref{action} as well as the equations of motion \eqref{eq:eom} still remain finite in the zero-curvature limit, and the equations of motion for background fields and perturbations in this limit should agree with ones where the spatial curvature can be ignored.

\subsection{Background and linear perturbations}
In the present paper, we adopt the Newtonian gauge whose metric is given by
\begin{align}
    g_{\mu\nu} dx^\mu dx^\nu
    =& -\left(1+2\Phi(t,\bf{x})\right) dt^2 
    \nn
    &+ a(t)^2 \left(1+2\Psi(t,\bf{x})\right) \delta_{ij} d{\bf x}^i d{\bf x}^j 
    \,.\label{eq:linearlineelement}
\end{align}
We define the St\"{u}ckelberg perturbations through the coordinate transformation,
\begin{align}
    x^\mu \to x^\mu + \delta^\mu_a \Pi^a
    \,,
\end{align}
and then the St\"{u}ckelberg fields become
\begin{align}
    \phi^{0} &=\frac{\alpha}{\sqrt{\kappa}}+\chi\left(t+\Pi^{0}\right)+\mathcal{O}(\sqrt{\kappa}) 
    \,,\\
    \phi^{i} &=\alpha\left(x^{i}+\partial^{i} \Pi\right)+\mathcal{O}(\sqrt{\kappa})
    \,.
\end{align}
Consequently, we obtain the perturbed $\bar{f}$-metric in the zero-curvature limit, which is given by
\begin{align}
    &\lim_{\kappa \to 0}  \bar{f}_{\mu\nu} d x^\mu dx ^\nu 
    \nn
    &~= \alpha^2 \partial^i \dot{\Pi} \partial_i \dot{\Pi} dt^2 + 2 \alpha^2(\partial_i\dot{\Pi}+\partial^k\dot{\Pi}\partial_k\partial_i\Pi) dt dx^i 
    \nn
    &~~~+ \alpha^2 (\delta_{ij}+2\partial_i\partial_j\Pi + \partial_i\partial^k\Pi\partial_k\partial_j\Pi) dx^i dx^j\,.
\end{align}
The matter perturbation is provided by $\rho(t,{\bf x}) = \rho(t) +\delta \rho(t,{\bf x})$, which gives
 \begin{align}
    &T^{({\rm m})0}{}_{~0} = -(\rho + \delta\rho)
    \,,\\
    &T^{({\rm m})0}{}_{~i} = 
    T^{({\rm m})i}{}_{~0} = 
    T^{({\rm m})i}{}_{~j} = 0
    \,.
\end{align}

\subsection{Background and linear equations}
By substituting \eqref{eq:g_zcsl} and \eqref{eq:f_zcsl} into \eqref{eq:eom} and \eqref{eq:matter_equation}, we obtain the background equations,
\begin{align}
    H^2 - \frac{\rho+\rho_g}{3\mpl^2} &=0
    \,,\label{backgroundF}\\
    \dot{H} + \frac{\rho+\rho_g+p_g}{2\mpl^2} &= 0
    \,,\\
    \dot{\rho}+3 H \rho &= 0
    \,,
    \label{background}
\end{align}
where we have defined the effective energy density and pressure
of the graviton mass term as
\begin{align}
    \rho_g &:= -{1\over 2}m^2\mpl^2 U\,,
    \label{eq:rhog}
    \\
    p_g &:={1\over 2}m^2\mpl^2 \Big(U- 2 \xi^2 U_{[Z]}-4\xi^4 U_{[Z^2]}-6\xi^6 U_{[Z^3]}\Big)
    \,,\label{eq:pg}
\end{align}
and $\xi:=\alpha/a$. Similarly, from \eqref{eq:Stuckelberg_equation}, we obtain the background St\"{u}ckelberg field equation,
\begin{align}
    \dot{\rho}_g+3 H(\rho_g+p_g) &= 0
    \,.
    \label{eq:St}
\end{align}
At the linear level, $(0,0)$ and traceless components of the Einstein equation \eqref{eq:eom} give
\begin{align}
    \frac{2}{a^2} \partial^2\Psi+6 H \left(H \Phi -\dot{\Psi} \right)~~~~~~~~~~~&
    \nn
    ~~~~~+\frac{(\rho_g+p_g) \left(\partial^2 \Pi-3 \Psi \right)}{\mpl^2}+\frac{\delta \rho}{\mpl^2}&=0
    \,,
    \label{eq:linear_00}
    \\
    \partial^2(\Phi+\Psi +a^2 M_{\rm GW}^2 \Pi)&=0
    \,,
    \label{eq:linear_tl}
\end{align}
where we define the tensor graviton mass as
\begin{align}
    M_{\rm GW}^2 =
    {2 \over \mpl^2} \Big[
    \rho_g + p_g - 2m^2 \mpl^2 \xi^4 
    \left(
        U_{[Z^2]}+3\xi^2 U_{[Z^3]} 
    \right)
    \Big]\,.
\end{align}
The time and space components of the linearized St\"{u}ckelberg equations are given by
\begin{align}
    \calC(-3\Psi+\partial^2\Pi+3H\Pi^0) &= 0\,, 
    \label{eq:Pi0}
    \\
   \partial_i\Big[2H\mpl^2M_{\rm GW}^2\partial^2\Pi-\dot{p}_g(\partial^2\Pi-3\Psi)~~~~~~~~~&
   \nn
   +3H(\rho_g+p_g)\Phi-\calC (\partial^2\Pi+3H\Pi^0-3\Psi)&
   \nn
    +a^2 H (2\dot\rho_g - 3\dot p_g) \dot \Pi -3 a^2 H (\rho_g +p_g) \ddot \Pi \Big]&=0\,,
    \label{eq:Pi}
\end{align}
where 
\begin{align}
    \calC &:= m^2 \mpl^2 \dot{\chi}^2 U_{\chi\chi}
    \,.
\end{align}
Integrating out $\Pi^0$ by using \eqref{eq:Pi0}, the equation for $\Pi$, \eqref{eq:Pi}, becomes
\begin{align}
   &2H\mpl^2M_{\rm GW}^2\partial^2\Pi-\dot{p}_g(\partial^2\Pi-3\Psi)
   +3H(\rho_g+p_g)\Phi
   \nn
   &~~+a^2 H (2\dot\rho_g - 3\dot p_g) \dot \Pi -3 a^2 H (\rho_g +p_g) \ddot \Pi = 0\,.
   \label{eq:reduced_Stuckelberg}
\end{align}

\subsection{Subhorizon expansion}
\label{Subsec:subhorizon}
Hereafter, we consider the spherically symmetric perturbations: ${\cal E}(t,\bm{x}) \to {\cal E}(t,r)$, where ${\cal E}=\{\Phi,\Psi,\Pi,\Pi^0\}$ and $r=\sqrt{x^2+y^2+z^2}$, and we use the following notation
\begin{align}
    \partial_i=\frac{x_i}{r}\frac{\partial}{\partial r}\,,
    \label{eq:spherical_derivative}
\end{align}
for simplicity. Now, we would like to derive equations applicable for subhorizon scales. To this end, we perform subhorizon expansion assuming that the time scale of the evolution of perturbations is roughly the Hubble time, i.e., $\dot{{\cal E}}\sim H{\cal E}$, and we also expand the perturbations as ${\cal E}= {\cal E}^{(1)} + {\cal E}^{(2)} + \cdots$, where ${\cal E}^{(n)} = \calO(\epsilon^n)$ and $\epsilon$ is a small expansion parameter. In order to extract the relevant terms in the subhorizon expansion, we additionally assume~\cite{Kenna-Allison:2020egn}
\begin{align}
    &r=\calO(\epsilon^{1/2})\,,
    \quad
    \del_i=\calO(\epsilon^{-1/2})\,.
\end{align}
This leads that the second spatial derivative of the perturbations at the leading order are the same order of the background, that is, $\partial^2 {\cal E}^{(1)} = \calO(\epsilon^0)$, and we therefore need to impose $\delta\rho=\calO(\epsilon^0)$ in order to reproduce the standard Poisson equation at the linear level, that is, $\delta\rho\sim a^{-2}\mpl^2 \del^2\Psi^{(1)}$.

Based on these assumptions, the linear equations \eqref{eq:linear_00}, \eqref{eq:linear_tl} and \eqref{eq:reduced_Stuckelberg}, of order $\calO(\epsilon^0)$ are respectively given by
\begin{align}
    \frac{2}{a^2}\del^2\Psi^{(1)}+\frac{\delta\rho}{\mpl^2}+\frac{\rho_g+p_g}{\mpl^2}\del^2\Pi^{(1)} &= 0\,,
    \\
    \del^2\Psi^{(1)}+\del^2\Phi^{(1)}+a^2\mgw^2\del^2\Pi^{(1)} &= 0\,,
    \\
    \del^2 \Pi^{(1)} &= 0\,,
\end{align}
where we have eliminated the background contributions by using \eqref{backgroundF}-\eqref{background} and \eqref{eq:St}.\footnote{The background equations can be safely used at this point since the solutions for the perturbations ${\cal E}^{(1)}$ are sufficiently small compared with the background as one can see from \eqref{eq:linear_sol_0}.} Imposing that the perturbations vanish at $r\to\infty$, the solution of these equations is given by
\begin{align}
    \Pi^{(1)} = \frac{Q(t)}{r}\,,\quad
    \Psi^{(1)} = \frac{G\calM}{ar}\,,\quad
    \Phi^{(1)} = -\frac{G\calM}{ar}\,,
    \label{eq:linear_sol_0}
\end{align}
where $\calM$ is an enclosed mass defined by
\begin{align}
    \calM(t,r):=4\pi a^3\int_0^r dr r^2 \delta \rho(r)\,,
\end{align}
and $Q$ is the integration constant. 
The gravitational potentials in \eqref{eq:linear_sol_0} are identical to ones in general relativity, and the fifth force due to the scalar graviton is absent at the leading order in the subhorizon expansion.

Equations for the next order perturbations  $\partial^2 {\cal E}^{(2)} = {\cal O}(\epsilon^1)$ are given by
\begin{align}
    &\frac{2}{a^2}\del^2\Psi^{(2)}+6H(H\Phi^{(1)}-\dot{\Psi}^{(1)})
    \nn
    &\qquad
    +\frac{\rho_g+p_g}{\mpl^2}(\del^2\Pi^{(2)}-3\Psi^{(1)}) = 0\,,
    \\ &\del^2\left(\Psi^{(2)}+\Phi^{(2)}+a^2\mgw^2\Pi^{(2)}\right) = 0\,,
    \\
    &2H\mpl^2M_{\rm GW}^2\partial^2\Pi^{(2)}-\dot{p}_g(\partial^2\Pi^{(2)}-3\Psi^{(1)})
    \nn    
    &\qquad+3H(\rho_g+p_g)\Phi^{(1)}+a^2 H (2\dot\rho_g - 3\dot p_g) \dot \Pi^{(1)}
    \nn
    &\qquad-3 a^2 H (\rho_g +p_g) \ddot \Pi^{(1)} = 0\,.
\end{align}
The solution of these equations can be obtained recursively. From \eqref{eq:linear_sol_0}, we get
\begin{align}
    \del^2\Pi^{(2)} =& \frac{3[H(\rho_g+p_g)-\dot p_g]}{2H \mpl^2\mgw^2-\dot p_g}\frac{G\calM}{ar}
    \nn
    &+(\text{terms~with~} \dot Q,\ddot Q)\,,
    \label{eq:pi_second}
    \\
    \del^2\Psi^{(2)} =& \frac{3a^2H}{2}\Bigg[\frac{\rho_g+p_g}{\mpl^2}\frac{2\mpl^2\mgw^2-(\rho_g+p_g)}{2H\mpl^2\mgw^2-\dot p_g}+6H\Bigg]
    \nn
    &\times\frac{G\calM}{ar}+(\text{terms~with~} \dot Q,\ddot Q)\,,    \label{eq:psi_second}
    \\
    \del^2\Phi^{(2)}
    =&-3a^2\Bigg[3H^2 +\mgw^2\frac{H(\rho_g+p_g)-\dot p_g}{2H\mpl^2\mgw^2 -\dot p_g}
    \nn
    &+\frac{H(\rho_g+p_g)}{2\mpl^2}\frac{2\mpl^2\mgw^2-(\rho_g+p_g)}{2H\mpl^2\mgw^2 -\dot p_g}
    \Bigg]\frac{G\calM}{ar}
    \nn
    &+(\text{terms with }\dot{Q},\ddot{Q})\,.
\label{eq:phi_second}
\end{align}
We separated $Q$-dependent terms, which are not explicitly written here. As we will see later, the matching condition to the nonlinear solution leads to $Q=0$, and thus $Q$-dependent terms vanish. Note that the solution for $\Pi^{(2)}$ completely agrees with the result in~\cite{Manita:2021qun}, and one immediately notice that 
$\Pi^{(2)}/\Pi^{(1)}\sim \Psi^{(2)}/\Psi^{(1)} \sim \Phi^{(2)}/\Phi^{(1)}\sim r/(aH)^{-1}$, which is extremely small inside the horizon.

\section{Nonlinear perturbations}
\label{spherical}
In this section, we investigate 
the effect of the nonlinearity of the additional scalar mode at small scales.
The basic idea is similar to the case of the Vainshtein mechanism since the mass potential in the action \eqref{action} yields higher-order (spatial) derivative interactions such as galileon theories. However, due to the presence of $[Z]$, $[Z^2]$, and $[Z^3]$ in an arbitrary function of the mass potential, the standard perturbative analysis leads to the infinite number of nonlinear terms responsible for the Vainshtein screening. 
To evade this problem, we, hereafter, consider a general polynomial up to quadratic order for $Z$ as the mass potential:
\begin{align}
    &U(X,[Z],[Z^2],[Z^3]) 
    \nn
    &= a_0(X) + a_1(X) [Z] + a_2(X) [Z]^2 + a_3(X) [Z^2] \,.
\end{align}

\subsection{Leading order}
\label{sec:leading_order}

In order to extract relevant nonlinear terms such as $(\partial^2 {\cal E})^n$, we again use the similar assumption introduced in Sec.~\ref{Subsec:subhorizon}, that is,
\begin{align}
    &{\cal E} =
    \{\Phi,\Psi,\Pi,\Pi^0\}=\calO(\epsilon), \notag\\   
    & \partial_i=\calO(\epsilon^{-1/2}),
    \quad \delta\rho=\calO(1)\,,
    \label{eq:counting_ansatz}
\end{align}
together with $\dot{{\cal E}}\sim H{\cal E}$ and the subhorizon limit, $r\ll (aH)^{-1}$ without expanding ${\cal E}$ itself. Note that $\calE$ used hereafter in this subsection corresponds to $\calE^{(1)}$ in the Sec.~\ref{Subsec:subhorizon}. Then, $\calO(1)$ equations contain all the nonlinear terms with $(\partial^2 {\cal E})^n$ including the background and linear contributions. Similarly to the linear case, we assume the background equations \eqref{backgroundF}-\eqref{background} and \eqref{eq:St} are satisfied, which can be easily verified after deriving nonlinear solutions. These equations are shown explicitly in Appendix~\ref{sec:nonlinear_equations}.

The St\"{u}ckelberg field equation can be easily obtained by combining \eqref{eq:stuckelberg_nl_0} and \eqref{eq:stuckelberg_nl_i} and eliminating the third spatial derivative $\Pi'''$,
\begin{align}
    (1+\Pi'')\calP_1\calP_2=0\,,
    \label{eq:master}
\end{align}
where
\begin{align}
    \calP_1 &= \calC_1 +\calC_2\left[\left(\frac{\Pi'}{r}\right)^2+2\frac{\Pi'}{r}\right]
    +\calC_3\left[(\Pi'')^2+2\Pi''\right]\,,
\end{align}
\begin{align}
    \calP_2 =~& \calC_4\left(\Pi''+2\frac{\Pi'}{r}\right) + 4\calC_5\left(\frac{\Pi'}{r}\right)^2
    \nn
    &-2\left(\frac{5\calC_4}{4}-2\calC_5\right)(\Pi'')^2
    \nn
    &-\left(\frac{\calC_4}{4}-\calC_5\right)\left[\left(\frac{\Pi'}{r}\right)^4+4\left(\frac{\Pi'}{r}\right)^3\right]
    \nn
    &-\left(\frac{3\calC_4}{4}-\calC_5\right)\left[(\Pi'')^4 +4 (\Pi'')^3\right]
    \nn
    &+\left(\frac{5\calC_4}{2}-2\calC_5\right)\left[(\Pi'')^2+2(\Pi'')\right]\left[\left(\frac{\Pi'}{r}\right)^2+2\frac{\Pi'}{r}\right]\,. 
\end{align}
We have utilized the symbol $'$ to represent the derivative with respect to $r$. We have also defined $\calC_n$ as
\begin{align}
    \calC_1 &= 6\mpl^2 \mgw^2 +\rho_g + 9p_g\,,
    \\
    \calC_2 &= -3\mpl^2\mgw^2+10\rho_g+18p_g\,,
    \\
    \calC_3 &= 
    3(3\mpl^2\mgw^2-4\rho_g)\,,
    \\
    \calC_4 &=12 \left[H(\rho_g+9p_g)+3\dot{p}_g\right]\,,
    \\
    \calC_5 &= H(12\mpl^2\mgw^2+5\rho_g+93p_g)
    \nn
    &~~+6\mpl^2\mgw\dot{M}_{\rm GW}+27\dot{p}_g\,.
\end{align}

As boundary conditions, we impose that
\begin{itemize}
    \item the solution is regular at $r=0$ 
    \item the solution obtained from \eqref{eq:master} connects to the linear solution \eqref{eq:linear_sol_0} at large distances.
\end{itemize}
Note that the first condition is imposed since
the St\"{u}ckelberg equations \eqref{eq:stuckelberg_nl_0} and \eqref{eq:stuckelberg_nl_i} have no source term. 
Manifestly, one of the solutions of \eqref{eq:master} given by $1+\Pi''=0$ does not satisfy the regularity condition, and we therefore disregard this solution. To see other solutions, we first expand the scalar graviton $\Pi'$ around $r=0$ as
\begin{align}
    \Pi'=\sum_{n=0}^\infty \alpha_n r^n\,.
\end{align}

In the case of $\calP_1=0$,
all coefficients except $\alpha_1$ are zero, thus there is an exact solution proportional to $r$, implying that this solution never matches the linear solution \eqref{eq:linear_sol_0}.
In addition,
the St\"{u}ckelberg field equation \eqref{eq:stuckelberg_nl_i} never hold for this solution unless $6\mpl^2\mgw^2+\rho_g+9p_g=0$, which cannot be realized in a general cosmological background. Therefore, this is not a physical solution, and we also disregard this solution.

As for $\calP_2=0$, only $\alpha_1$ is the non-vanishing coefficient, and there are three branches of solutions: 
\begin{align}
    \Pi'=
    \begin{cases}
    -r\,,
    \\
    -2r\,,
    \\
    0\,,
    \end{cases}
\end{align}
which are also the exact solutions of $\calP_2=0$ as well as \eqref{eq:stuckelberg_nl_0} and \eqref{eq:stuckelberg_nl_i}. The first two solutions proportional to $r$ does not satisfy the matching condition with the linear solution \eqref{eq:linear_sol_0} at large scales, and we again disregard these solutions. Therefore, $\Pi'=0$ is the only solution satisfying all the conditions, and the matching condition leads to the vanishing of the integration constant in the linear solution of $\Pi'$ given in \eqref{eq:linear_sol_0}, that is,
\ba
Q=0 \,.
\ea
This indicates that the scalar graviton never appears even at small scales in the leading order of the quasi-static approximation. Therefore, we proceed to investigate the subleading order equations.

\subsection{Subleading order}
In the previous subsection, we 
extracted
the relevant nonlinear terms from the equation of motion under the assumption~\eqref{eq:counting_ansatz},
but this results in $\Pi'=0$ for all scales within the horizon. Thus, as in the case of the linear perturbations, we need to further expand the perturbations as ${\cal E}= {\cal E}^{(1)} + {\cal E}^{(2)} + \cdots$ with ${\cal E}^{(n)} = \calO(\epsilon^n)$. Note that the analysis done in the previous subsection is for $\calE^{(1)}$. Then, all the equations of motion for $\calE^{(2)}$ completely coincide with ones in the linear perturbations, i.e., the solutions are simply given by~\eqref{eq:pi_second}-\eqref{eq:phi_second}. 
In fact, the scalar graviton completely decouples with the metric perturbations $\Psi$ and $\Phi$ in the decoupling limit, and thus the gravitational potentials are exactly the same as ones in Newtonian case, whose results are shown in the Appendix.~\ref{sec:decoupling_limit}.

\section{Summary}
\label{discussion}
In the present paper, we consider both linear and nonlinear perturbations in a spherically symmetric configuration around the homogeneous and isotropic background in PMG. In order to simplify the analysis, we utilized the zero-curvature limit, which is applicable where the spatial curvature can be ignored. Having in mind that nonlinear derivative interactions appearing in the galileon-like theories are also present in PMG, we derived the nonlinear equations for all perturbations and found the analytical solutions. The condition for the nonlinear solution to be connected to the linear solution at large scales requires the vanishing of the integration constant for the leading solution of $\Pi$ defined in \eqref{eq:linear_sol_0}, that is, $Q=0$. This indicates that 
the scalar graviton
should be zero at the lowest order of the subhorizon expansion. 
As we can easily see in \eqref{eq:ModifiedEinsteinNL00} and \eqref{eq:ModifiedEinsteinNLtl}, the modified Einstein equations 
remains the same as in general relativity.

We have also confirmed that, even at the subleading order in the subhorizon expansion,  the nonlinear effect due to the scalar graviton is absent at all scales, and the linear contributions dominate at all scales, which are highly suppressed as can be seen in the Poisson equation.
In fact, 
one of the PPN parameters $\gamma:=-\Psi/\Phi$ can be roughly estimated as
\begin{align}
    1-\gamma &\sim \frac{a^2M_{\rm GW}^2\Pi^{(2)}}{\Phi^{(1)}}
    \sim r^2/(aH)^{-2}\,,
\end{align}
where we have assumed 
$\mgw \sim H$ and $\rho_g\sim p_g\sim \mpl^2H^2$.
This implies that the correction of $\gamma$ to general relativity is highly suppressed well inside the horizon.
For example, at the solar system scale, this is estimated as $1-\gamma\sim10^{-29}$, and is much smaller than the constraint by the Cassini spacecraft $1-\gamma \lesssim 10^{-5}$~\cite{Bertotti:2003rm}.

In summary, the effect of modification due to the scalar graviton is strongly suppressed in the subhorizon region. As we show in the Appendix \ref{sec:decoupling_limit}, 
this result is consistent with the decoupling limit analysis, and 
the scalar mode 
interacts with neither the tensor mode and external matter fields.

\begin{acknowledgements}
We thank Tsutomu Kobayashi and Takahiro Tanaka for insightful comments. Y.M. acknowledges the xPand package \cite{Pitrou:2013hga} that was used to explicitly confirm calculations. 
This work is supported by
the establishment of university fellowships towards the creation of science technology innovation (Y.M.), Japan Society for the Promotion of Science (JSPS) Overseas Challenge Program for Young Researchers (Y.M.), JSPS Grants-in-Aid for Scientific Research No. JP22K03605 (R.K.), and Waseda University Grant for Special Research Project No. 2022C-632 (S.P.).
\end{acknowledgements}

\appendix

\section{Nonlinear equations at leading order}
\label{sec:nonlinear_equations}

In this section, we show the leading-order nonlinear equations derived under the assumption \eqref{eq:counting_ansatz} in a spherically symmetric setup. 
With the order ansatz \eqref{eq:counting_ansatz}, we extract all $\calO(\epsilon^0)$
terms from the Einstein and St\"{u}ckelberg equations, which are given by
\begin{align}
    \calE^{00}=&-\frac{2}{a^2}\del^2\Psi-\frac{\delta\rho}{\mpl^2}-\frac{3\calC_1-2\calC_3}{27\mpl^2}[\Pi]-\frac{\calC_2}{18\mpl^2}[\Pi]^2
    \nn
    &-\frac{\calC_1-\calC_2}{18\mpl^2}[\Pi^2]+\frac{3\calC_2-2\calC_3}{54\mpl^2}\left([\Pi^3]+\frac{1}{4}[\Pi^4]\right)
    \nn
    &-\frac{\calC_2}{18\mpl^2}\left([\Pi][\Pi^2]+\frac{1}{4}[\Pi^2]^2\right)\,,
    \label{eq:ModifiedEinsteinNL00}
    \\
    \calE^{\rm tl} =& \frac{1}{a^2}\del^2(\Phi+\Psi)+\frac{6\calC_1-3\calC_2-2\calC_3}{27\mpl^2}[\Pi]
    \nn
    &+\frac{\calC_1-4\calC_2+\calC_3}{9\mpl^2}[\Pi^2]+\frac{\calC_2}{18\mpl^2}(4[\Pi]^2+3[\Pi^2])
    \nn
    &-\frac{2(\calC_2-2\calC_3)}{27\mpl^2}\left([\Pi^3]+\frac{1}{4}[\Pi^4]\right)
    \nn
    &+\frac{2\calC_2}{9\mpl^2}\left([\Pi][\Pi^2]+\frac{1}{4}[\Pi^3]\right)\,,
    \label{eq:ModifiedEinsteinNLtl}\\
        \calE^{\rm St,0} =& -2\calC_4[\Pi]
    +2(5\calC_4-6\calC_5)[\Pi^2]
    \nn
    &-(5\calC_4-4\calC_5)\left([\Pi]^2+[\Pi^2][\Pi]+\frac{1}{4}[\Pi^2]^2\right)
    \nn
    &+(11\calC_4-12\calC_5)\left([\Pi^3]+\frac{1}{4}[\Pi^4]\right)
    \nn
    &+\frac4 3\calE^{{\rm St},i}\dot{\Pi}_i(1+\hat{x}^k\hat{x}^l\Pi_{kl})^{-1}
    \,,
    \label{eq:stuckelberg_nl_0}
    \\
    \calE^{{\rm St},i} =&6\calC_1 \Pi^{ij}{}_{j} + 3(2\calC_1-\calC_2+2\calC_3)\Pi^i{}_j \Pi^{jk}{}_{k}
    \nn
    &-3(\calC_2-2\calC_3)\Pi^{jk}\Pi^{i}{}_{jk}+8\calC_3\Pi^{ij}\Pi^{kl}\Pi_{jkl}
    \nn
    &-(3\calC_2-2\calC_3)\Big((\Pi^2)_{jk}\Pi^{ijk}+\Pi^{ij}(\Pi^2)^{kl}\Pi_{jkl}
    \nn
    &+(\Pi^3)^{i}{}_j\Pi^{jk}{}_{k}+(\Pi^2)^{ij}\Pi^{kl}\Pi_{jkl}\Big)\nn
    &-2(3\calC_2-4\calC_3)(\Pi^2)^{i}{}_j\Pi^{jk}{}_{k}
    \nn
    &+3\calC_2\Big(2[\Pi]\Pi^{ij}{}_{j}+2\Pi^{ij}[\Pi]\Pi_{jk}{}^{k}+[\Pi^2]\Pi^{ij}{}_{j}
    \nn
    &+\Pi^{ij}[\Pi^2]\Pi^{k}{}_{kj}+2(\Pi^2)^{ij}\Pi^{kl}\Pi_{jkl}\Big)\,,
    \label{eq:stuckelberg_nl_i}
\end{align}
where $\hat{x}^i$ 
is defined by $\hat{x}^i:=x^i/r$, $\Pi_i=\del_i \Pi$, $\Pi_{ij}=\del_i \del_j\Pi$, $(\Pi^2)_{ij} = \Pi_{ik}\Pi^k_{~j}$, and so on. 
Note that the St\"{u}ckelberg equations, \eqref{eq:stuckelberg_nl_0} and \eqref{eq:stuckelberg_nl_i}, does not contain the gravitational potentials $\Psi$ and $\Phi$ and the source term. 

\section{Decoupling limit}
\label{sec:decoupling_limit}
In this appendix, we derive the effective action in the so-called decoupling limit~\cite{Luty:2003vm} to see the role of the scalar graviton. Hereafter, we consider $G=1$. 

In order to take the decoupling limit, we, for simplicity, utilize the zero-curvature scaling limit defined through \eqref{scaling}. Then the effective action in the zero-curvature limit is provided by
\ba
S
&=&
\int {\rm d}^4 x \sqrt{-g} \frac{\mpl^2}{2}\Biggl[ \,R   
+ m^2 U\big({\cal X}, [\calZ] ,[\calZ^2] ,[\calZ]^2 \big) \Biggr] \notag\\
&&+ S_{\rm m}  [g, \psi] \,,
\label{action: effective}
\ea
where ${\cal X} = {\tilde \phi}^a {\tilde \phi}_a$ and $\calZ =  g^{\mu\alpha} \delta_{ij}^{(3)} \partial_\alpha {\tilde \phi}^i\partial_\nu{\tilde \phi}^j$. Here, the indices $i$ and $j$ are the spacial indices and $\delta_{ij}^{(3)}$ is the Kronecker delta in 3D space. In a flat FLRW metric, one can easily revive the same background equations \eqref{backgroundF}-\eqref{background} with the St\"{u}ckelberg configuration, ${\tilde \phi}^0  = \chi(t)$ and $ {\tilde \phi}^i = x^i$, from the action \eqref{action: effective}. 
In the following, we choose the mass potential as 
\ba
U= a_1({\cal X})  [\calZ] +a_2({\cal X})  [\calZ]^2 + a_3({\cal X}) [\calZ^2] \,.
\ea

Now we decompose the St\"{u}ckelberg field ${\tilde \phi}^a$ around unitary gauge, ${\tilde \phi}^a=\delta^{a}_{~\mu} x^\mu$, 
\ba
{\tilde \phi}^a = x^a - V^a - \partial^a \varphi\,,
\ea
into the vector mode $V^a$ and the scalar mode $\varphi^a$, and we perturb the fluctuation of the metric around a Minkowski metric, $g_{\mu\nu} = \eta_{\mu\nu} + h_{\mu\nu}/\mpl$. 
Then we take the decoupling limit noting that
\ba
m \to 0, \qquad \mpl^2 \to \infty, 
\ea
keeping 
\ba
\sqrt{m \mpl}  \equiv \Lambda_2 \to {\rm fixed}, \qquad 
T_{(\rm m)}^{\mu\nu}}/\mpl \to {\rm fixed\,.
\ea
The effective Lagrangian in the decoupling limit describes physics at distances in the range $\Lambda_2^{-1} < r  < m^{-1} $. Since $m\sim H_0$ in order to explain the accelerated expansion of the universe at present~\cite{Manita:2021qun}, the decoupling limit approach overlaps the analysis using the subhorizon approximation introduced in Sec.~\ref{PMG} and~\ref{spherical}. 
Then the decoupling limit Lagrangian ignoring the vector mode is given by
\ba
{\cal L}_{\rm DL} &=&-{1\over 4} h^{\mu\nu} {\cal E}^{\alpha\beta}_{\mu\nu}h_{\alpha\beta} + h_{\mu\nu} 
T_{(\rm m)}^{\mu\nu}\notag\\
&&+a_1 \, \Lambda_2^4 \left[
3+ \delta_{ij}^{(3)} \left(
\p^\mu \p_i \varphi \p_\mu \p_j \varphi 
-2\p_i \p_j \varphi 
\right)
\right]\notag\\
&&+F_{\rm NL}(\Lambda_2, \partial_\mu \varphi, \partial_\mu \partial_i \varphi) \,,
\ea
where 
${\cal E}^{\mu\nu\alpha\beta}$ is the linearized Einstein-Hilbert kinetic operator defined as 
\ba
{\cal E}^{\mu\nu}_{~~\alpha\beta}
&=& 
\left[ \eta^{(\mu}_{~\alpha}\eta^{\nu)}_{~\beta}-\eta^{\mu\nu}\eta_{\alpha\beta}\right] \square
-2 \partial^{(\mu}\partial_{(\alpha}\eta^{\nu)}_{~\beta)}\notag\\
&&+\partial^\mu \partial^\nu \eta_{\alpha\beta} + \partial_\alpha \partial_\beta \eta^{\mu\nu} \,.
\ea
The decoupling limit Lagrangian does not contain higher-order time derivatives, i.e., each scalar mode $\varphi$ has up to first derivative with respect to time, and this clearly shows the absence of the BD ghost. 
Differently from the dRGT case, the cutoff energy scale is $\Lambda_2 > \Lambda_3 \equiv (m^2 \mpl)^{1/3}$. 
Furthermore, the nonlinear interactions $F_{\rm NL}$ depend on only $\varphi$ and its derivatives, which means the tensor mode and scalar mode are completely decoupled in the decoupling limit. As a result, the matter field $T_{(\rm m)}^{\mu\nu}$ simply couples to graviton only through $h_{\mu\nu}$, and the gravitational potential is given by the Newtonian one. For this reason, the vDVZ discontinuity is manifestly absent. This agrees with our results in the analysis obtained in  Sec.~\ref{spherical}.

\bibliography{ref}

\end{document}